\documentclass[preprint]{elsarticle}
\usepackage{graphicx}
\usepackage{lineno, hyperref}
\usepackage[english]{babel}
\usepackage{color}

\journal{Computer Physics Communications}

\begin{document}

\begin{frontmatter}

\title{Large-scale Monte Carlo simulations for the depinning transition in Ising-type lattice models}

\author[mymainaddress]{Lisha Si}

\author[mymainaddress]{Xiaoyun Liao}

\author[mymainaddress]{Nengji Zhou\corref{mycorrespondingauthor}}

\cortext[mycorrespondingauthor]{Corresponding author}
\ead{zhounengji@hznu.edu.cn}

\address[mymainaddress]{Department of Physics, Hangzhou Normal University, Hangzhou 310046, China}

\begin{abstract}
With the developed ``extended Monte Calro'' (EMC) algorithm, we have studied the depinning transition
in Ising-type lattice models by extensive numerical simulations, taking the random-field Ising
model with a driving field and the driven bond-diluted Ising model as examples. In comparison
with the usual Monte Carlo method, the EMC algorithm exhibits greater efficiency of the simulations.
Based on the short-time dynamic scaling form, both the transition field and critical exponents of the
depinning transition are determined accurately via the large-scale simulations with the lattice size up
to $L=8~912$, significantly refining the results in earlier literature. In the strong-disorder regime,
a new universality class of the Ising-type lattice model is unveiled with the exponents $\beta=0.304(5),
\nu=1.32(3), z=1.12(1)$, and $\zeta=0.90(1)$, quite different from that of the Edwards-Wilkinson equation.
\end{abstract}

\begin{keyword}
\texttt{Dynamic critical phenomena, Monte Carlo methods, Depinning transition, Domain-wall dynamics}
\end{keyword}

\end{frontmatter}


\section{Introduction}
Driven by a constant force in the presence of the quenched disorder, the
interface moves with a steady-state velocity, while it is pinning when
the force is weak compared to the random noise. Between them, there exists a
second-order dynamical phase transition, called as the ``{\it depinning
transition}'' \cite{lem98,bol04,dou06,kol06a,luo07,bak08,im09,
gor14}. For several decades, the depinining transition has been the
focus of the experimental and theoretical research, which are common
to a wide variety of phenomena, including the liquid invasion in
porous media \cite{ros07}, the contact line in wetting \cite{du14},
the vortices in type-II superconductors \cite{bla94, oku12}, the
charge-density waves \cite{bra04}, the fracture propagation
\cite{leb13, ati15}, the dislocation dynamics in crystal plasticity
\cite{ova15}, and the domain-wall motions in ferromagnetic  and
ferroelectric materials \cite{lee11, ban14, shi15, ram15}.
Practically, understanding the fundamental mechanism of the
depinning transition plays an important role in predicting and
controlling the motions of the magnetic domain walls in
nano-materials \cite{wut13, dol14, fuk15}, thin films
\cite{gor14,she15}, and semiconductors \cite{ram15, yam07}, which is
key to the realization of the new classes of potential nonvolatile
storage-class devices \cite{hay08,par08}.

Theoretical approaches to the domain-wall dynamics are typically
based on the phenomenological models, such as the Edwards-Wilkinson
equation with quenched disorder (QEW) and its variants \cite{fer13,
bol14, jag14, che15}. With these equations, the domain wall in a
two-dimensional system can be effectively described by a
single-valued elastic string, and the static and dynamic critical
exponents of the depinning transition, i.e., $\beta, \nu, z$, and
$\zeta$, are measured numerically, though the discrepancies are
still large in the literature \cite{lop97, kol06, kim06, due05}.
For example, it reaches nearly $30$ percent in the velocity exponent $\beta$.
Recently, extensive simulations of the QEW at
the depining transition have been performed with a lattice size up to $L=8~192$ \cite{fer13}.
Based on the short-time dynamics method, the universality class of the
depining transition is identified with the exponents
$\beta=0.245(6), z=1.433(7), \zeta=1.250(5)$, and $\nu=1.333(7)$,
which are robust under the changes of the disorder realization
including the random-bond and random-field characters. Moreover, the
scaling relation $\beta=\nu(z-\zeta)$ is revealed, consistent with
the prediction of the functional renormalization group theory \cite{nat92, dou02}.
However, most experiments reported that the roughness exponent is $\zeta
\approx 0.6-0.9$ \cite{lem98,ati15,che15,jos98,met07},
smaller than that of the QEW equation, suggesting
that detailed microscopic structures and interactions of real
materials should be concerned.

Besides, the dynamical behaviors of the domain walls in ferromagnetic
nanowires are also investigated via the micromagnetic simulations
with the Landau-Lifshitz-Gilbert (LLG) equation depicting the
time evolution of the orientation of the magnetization distribution,
$m(\vec{r},t)$ \cite{elt10, gol10, ger14}.
However, the LLG equation is too complicated to be
simulated for the depinning transitions in ultrathin ferromagnetic or
ferroelectric films. The Ising-type lattice models are then
introduced with much simpler microscopic structures and interactions \cite{now98,col06, koi10,xi15}.
In this paper, we use the random-field Ising model with a driving field
(DRFIM) and the driven bond-diluted Ising model (DBDIM) as examples. For
a long time, it has been invariably stated that the QEW equation and
DRFIM model belong to the same universality class \cite{now98, xi15, ama95}.
However, significant deviations of the critical exponents have been reported in
recent works \cite{zho09,zho10a}, which could not be ruled
out by statistical errors. It was argued that the difference may be
induced by the intrinsic anomalous scaling and spatial multiscaling
of the DRFIM at the depinning transition. Unfortunately, a weak
dependence of the critical exponents on the strengthes of the random
fields is found in the disorder regime $\Delta \in [0.8, 2.3]$ \cite{qin12}.
Hence, it remains ambiguous that the depinning transition of the
DRFIM belongs to a new dynamic universality class or it only has a
correction to the universality class of the QEW equation due to the
influence of the first-order phase transition occurring at $\Delta \leq 1$.
To solve this issue, we will identify the critical exponents of DRFIM
in the regime of the strong disorder $\Delta \gg 1$ in this article.

Early studies of the depinning transition were always focused on the
steady-state velocity $v(L)$ of the domain wall \cite{kol06a,due05, now98, rot99}.
Suffering from severe critical slowing down, however, it is quite arduous to obtain
the exact transition field $H_c$ and critical exponents. Adopting the
short-time scaling form \cite{zhe98,luo98}, both the static and dynamic exponents
$\beta, \nu, \zeta$, and $z$ can be easily and accurately determined from
the nonsteady relaxation of the domain interface since the spatial correlation length
is short \cite{ fer13, kol06, zho09}. Due to the limitation of the computing
resources, however, the system size and simulation time are insufficient in
previous work for the depinning transition in the DRFIM, which are up to
$L=1~024$ and $t_{max}=2~000$ \cite{zho09, qin12},
much smaller than those in the QEW equation, $L=8~192$ and
$t_{max}=8~000$ \cite{fer13}. It may result in a systematic error
in the determination of the critical driving field $H_c$.
Accordingly, larger spatial and temporal scales are
needed in the simulations to obtain more precise results for
the depining transition in the DRFIM.

In this paper, an optimized Monte Carlo method is developed, termed
as the ``extended Monte Calro'' (EMC) algorithm. Adopting the EMC,
much smaller time of the Central Processing Unit (CPU) is taken for
the depinning transition, in comparison with that of the usual Monte
Carlo method. By extensive simulations, the transition point and
critical exponents of the DRFIM are then accurately determined for
various strengthes of the quenched disorder, and a new universality
class is unveiled. In addition, the depinning transition in another
Ising-type lattice model, DBDIM, is also investigated for
comparison. In Sec. $2$, the models, EMC algorithm, and scaling
analysis are described, and in Sec. $3$, the numerical results are
presented. Finally, Sec. $4$ includes the conclusion.

\section{Methodology}
\subsection{Model}
The DRFIM is one of the simplest demonstration to study the depinning transition
in the disorder media with microscopic structures and interactions.
The Hamiltonian of the DRFIM can be written as
\begin{equation}
\mathcal{H} = - J \sum_{<ij>}S_iS_j - H \sum_i S_i - \sum_i h_i S_i,
\label{equ100}
\end{equation}
where $S_i = \pm 1$ is the classical Ising spin of the
two-dimensional rectangle lattice with $2L \times L$, the random field
$h_i$ is uniformly distributed within an interval $[-\Delta,
\Delta]$, and $H$ is a homogeneous driving field. The initial state
that spins are positive in the sublattice on the left side and
negative on the right side, is used to build a perfect domain wall,
also referred to as a ``domain interface'', in the $y$ direction. The
direction perpendicular to the domain interface is then set to the
$x$ axis. Antiperiodic and periodic boundary conditions are used in
$x$ and $y$ directions, respectively. To eliminate the pinning
effect irrelevant for disorder, we rotate the square lattice such
that the initial domain wall orients in the $(11)$ direction of the
square lattice, as shown in Refs. \cite{now98,zho09,rot99,rot01}.

\begin{figure}[tbp]
  \centering
   \includegraphics[clip, scale=0.3]{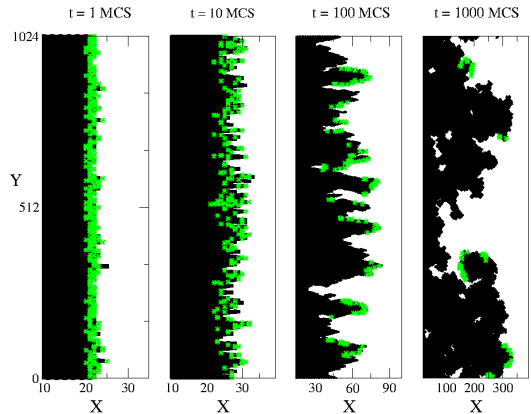}
   \caption{Time evolution of the spin configuration under the uniform
   distribution of the random fields. The black and white correspond to
   the spin up ($S_i = 1$) and down ($S_i =-1$), respectively, and the stars denote
   the activated spins within the domain interface. As the time $t$ grows,
   overhangs and islands are created. }
   \label{f100}
\end{figure}

After preparing the initial state, a usual Monte Carlo method is adopted with
standard one-spin flips in the simulations. Simply
speaking, we update each spin with the following procedure.
Firstly, we randomly choose a single spin $S_i$ in the
two-dimensional lattice. The change of the total energy is then
calculated after we flip the spin $S_i \rightarrow S_i'$,
\begin{eqnarray}
\delta E & = & \mathcal{H}(S_i) - \mathcal{H}(S_i')  \nonumber \\
          & = & (S_i'-S_i)\left[-J\sum_{j}S_j - H - h_i \right].
\label{equ110}
\end{eqnarray}
Only when $\delta E < 0$, the flip is accepted, otherwise the spin state
$S_i$ remains. A Monte Carlo time step (MCS) is defined by $2L^2$ single-spin updates.
As time evolves, the domain wall moves and roughens, while the bulk, i.e., spins far away
from the domain interface, keeps invariant. As shown in
Fig.~\ref{f100}, the time evolution of the spin configuration
is displayed with the black and whites squares, corresponding to
$S_i=\pm 1$, respectively. Complicated spin structure are found nearby the
domain interface, such as overhangs and islands at the time
$t=1~000$ MCS.

According to earlier literatures \cite{lem98, jos98, zho10a},
there are many different ways to define the domain interface.
In this paper, we adopt a simple and popular definition based on the magnetization.
Denoting a spin at site $(x, y)$ by $S_{xy}(t)$,
a microscopic height function of the domain
interface is introduced,
\begin{equation}
h(y,t) = \frac{L_x}{2}[ m(y,t) + 1], \label{equ120}
\end{equation}
where $m(y,t)$ is the linear magnetization defined as
\begin{equation}
m(y,t) = \frac{1}{L_x} \left[ \sum_{x=1}^{L_x} S_{xy}(t) \right],
\label{equ130}
\end{equation}
and $L_x$ is the size of the lattice in the $x$ direction.
The velocity of the domain interface is then calculated as
\begin{equation}
v(t) = \frac{d\langle h(y,t)\rangle}{dt}, \label{equ140}
\end{equation}
where $\langle \cdots \rangle$ includes both the statistical
average over samples and in the $y$ direction.
To depict the roughening process of the domain interface, the
roughness function $\omega(t)$ and correlation function
$C(r,t)$ are measured with
\begin{equation}
\omega^{(2)}(t) = \left \langle h(y,t)^2 \right \rangle - \langle
h(y,t) \rangle^2, \label{equ150}
\end{equation}
and
\begin{equation}
C(r, t) = \left\langle h(y + r, t)h(y, t) \right \rangle - \left\langle h(y,t) \right\rangle^2,
\label{equ160}
\end{equation}
respectively. The former describes the roughening of the domain
interface in the $x$ direction, and the latter reflects the growth of the spatial
correlation in the $y$ direction. Moreover, the function $F(t)$ is measured
as the ratio of the planar susceptibility and line susceptibility \cite{zho09,zho10a,qin12},
\begin{equation}
F(t) \sim [M^{(2)}(t) - M(t)^2]/\omega^2(t), \label{equ165}
\end{equation}
where $M(t)$ is the total magnetization, and $M^{(2)}(t)$ is its
second moment.

In addition, other types of the quenched disorder are also considered
for comparison, taking the random-bond disorder in the DBDIM as an example.
The Hamiltonian of the DBDIM is
\begin{equation}
\mathcal{H} = - \sum_{<ij>}J_{ij}S_iS_j - H \sum_i S_i,
\label{equ170}
\end{equation}
where $J_{ij}= 1 + \varepsilon_{ij}$ denotes the nearest-neighboring
coupling strength, and $\varepsilon_{ij}$ is the bond disorder following
a Gaussian distribution with the mean zero and the standard deviation
$\sigma$. Without loss of generality, we set $\sigma=1.5$ in the
simulations as an example of the strong-disorder case.

\subsection{Extended Monte Carlo (EMC) algorithm}
In usual Monte Carlo simulations, we are limited to the system size
$L_x=2L_y=2~048$ up to $t_{max}=2~000$ MCS for a sufficiently large
sample size $N_s=20~000$ \cite{zho09,qin12}, since all of the spins
in the lattice, on average,  should be chosen and updated in each MCS. However,
the number of the activated spins $N(t)$ decreases quickly with the time $t$,
as shown in Fig.~\ref{f100}. Where the activated spin is defined that each attempt of the flip
will be accepted once the spin is chosen. As an example, $N(t)=68$ is obtained at $t=1~000$ MCS, much smaller than
the total spin number $2L^2 \approx 10^6$. Consequently, most of the
spins are inactivated in Monte Carlo simulations. To avoid the undesirable waste of the time,
the EMC algorithm is developed in this paper with the architecture:

Step $1$: Initialize the spin lattice with a perfect domain
interface. The quenched disorder $h_i(x,y)$ is generated randomly
for each site in the rectangle lattice.

Step $2$: Create a table list of the activated spins. To be
specific, we search for the spins with the energy change $\delta E
<0$ according to Eq.~(\ref{equ110}), and store their
position information $(x,y)$ in the table list. The number of
activated spins $N(t)$ is then obtained as the size of the table list.

Step $3$: Update the table list. In this table list, we randomly choose a spin,
flip it with $S_i\rightarrow -S_i$, and delete its position information.
Afterwards, the change of the energy $\delta E$ is respectively calculated for its four nearest-neighboring spins.
If $\delta E <0$, new position information is added into the table list.

Step $4$: Increase the time of the EMC simulation to $t'=t+1/N(t)$ MCS.
Physical observable of the system including the microscopic
height function $h(y,t')$, roughness function $\omega(t')$, and
correlation function $C(r,t')$), are then assessed.

Step $5$: Repeat the steps $3$ and $4$ until the time of the
simulation $t>t_{max}$ indicating that a sample of the simulation
is terminated.

Step $6$: Repeat the steps $1 \sim 5$ for the statistical average
over samples and realizations of the quenched disorder.
The velocity of the domain interface $v(t)$ is then calculated
with Eq.~(\ref{equ140}).

With the EMC algorithm, the depinning transition with lattice sizes from
$L=32$ to $12~000$ is investigated up to $t_{max}=22~000$ MCS. Our
main results are presented at $L=8~192$, much larger than $L=1~024$
in earlier literatures \cite{zho09,qin12}, and simulation results at $L=12~000$
confirm that finite-size effects are already negligibly small.
Besides, the influence of the disorder strength is also
investigated, taking the uniformly distributed random field $h_i$
with the strength varying from $\Delta=0$ to $10$. For each set of
the parameters, more than $20~000$ samples are performed for
average. Statistical errors are estimated by dividing the total
samples in three subgroups. If the fluctuation in the time direction
is comparable with or larger than the statistical error, it also
will be taken into account.

\subsection{Scaling analysis}
As the depinning transition is of second order, the short-time scaling theory \cite{zhe98,luo98}
is applicative for the time evolution of the order parameter $v(t)$.
The dynamic scaling form is derived by scaling arguments with
a finite lattice size $L$ and nonequilibrium
spatial correlation length $\xi \sim t^{1/z}$,
\begin{equation}
v(t, \tau, L) = b^{-\beta/\nu} G(b^{-z}t, b^{1/\nu}\tau,
b^{-1}L), \label{equ180}
\end{equation}
where $b$ denotes an arbitrary rescaling factor, $\beta$ and $\nu$
correspond to the static exponents, $z$ is the dynamic exponent, and
$\tau=(H-H_c)/H_c$. Setting $b \sim t^{1/z}$, the scaling form can
be simplified in the short-time scaling regime with $\xi(t)\sim
t^{1/z} \ll L$,
\begin{equation}
v(t, \tau) = t^{-\beta / \nu z}G(t^{1/\nu z}\tau).
\label{equ190}
\end{equation}
Only at the transition point $\tau = 0$, a power law behavior is
expected,
\begin{equation}
v(t) \sim t^{-\beta / \nu z}. \label{equ200}
\end{equation}
The critical field $H_c$ is then located by searching for the best
power-law behavior of $v_{_M}(t, \tau)$. Afterwards, the critical exponent $\beta /\nu z$
is estimated from Eq.~(\ref{equ200}), and $1/\nu z$ is measured from the time
derivative of $v(t, \tau)$ in Eq.~(\ref{equ190}),
\begin{equation}
\left.\frac{\partial \ln v(t, \tau)}{\partial \tau}
\right|_{\tau=0}\sim t^{1/\nu z}. \label{equ210}
\end{equation}

The nonequilibrium correlation length $\xi(t)$ can also be extracted
independently from the correlation function $C(r,t)$ defined in
Eq.~(\ref{equ160}) with the scaling form
\begin{equation}
C(r,t) = \omega^{2}(t) \widetilde{C}\left( r/\xi(t) \right),
\label{equ220}
\end{equation}
where $\widetilde{C}(s)$ is the scaling function with $s=r/\xi(t)$,
and $\omega^{2}(t)$ is the roughness function defined in
Eq.~(\ref{equ150}). With the correlation length $\xi(t)$ at hand,
the roughness exponent $\zeta$ is estimated from the kinetic
roughening of the domain interface,
\begin{equation}
\omega^{2}(t) \sim \left[\xi(t)\right]^{2\zeta}. \label{equ230}
\end{equation}
Meanwhile, one may determine the local roughness exponent
$\zeta_{loc}$ by fitting $\widetilde{C}(r, t)$ with an empirical
scaling form,
\begin{equation}
\widetilde{C}(r/\xi(t)) \sim \exp\left[-\left( r /
\xi(t)\right)^{2\zeta_{loc}}\right]. \label{equ240}
\end{equation}

Though power law behaviors at the critical point are expected in
Eqs.~(\ref{equ200}),(\ref{equ210}) and (\ref{equ230}) based on the scaling theory,
corrections to scaling should be considered to extend the fitting to the early times.
Usually, a power-law correction form is adopted,
\begin{equation}
y = ax^b(1+c/x), \label{equ250}
\end{equation}
where the fitting parameter $b$ corresponds to the critical exponent.

\section{Numerical results}
\subsection{Superiority of the EMC algorithm}
As it is referred above, the usual Monte Carlo method is time consuming
to study the depinning phase transition in disordered media.
To overcome it, an extensive version of the Monte Carlo
method, i.e., EMC algorithm, has been introduced. In
this subsection, we will compare the results obtained from the EMC
with those obtained from the usual Monte Carlo method, including the
CPU time and the velocity $v(t)$ of the domain interface.

\begin{figure}[tbp]
  \centering
   \includegraphics[clip, scale=0.3]{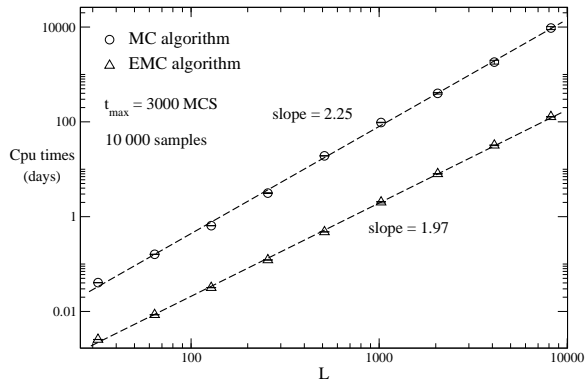}
   \caption{ CPU time of the simulations based on the usual Monte Carlo algorithm (open circles) and EMC algorithm
   (open triangles) are displayed as a function of the lattice size $L$. The unit of the CPU time is one day.
   For each case of $L$, $10~000$ statistical samples are performed for average, up to $t_{max}=3000$ MCS,
   and error bars are also given. Dashed lines represent power-law fits.
   }
   \label{f110}
\end{figure}

In Fig.~\ref{f110}, the CPU time are shown for these two algorithms.
For convenience of the comparison, we set the maximum value of the
simulation time $t_{max}=3~000$ MCS and the size of the statistical
samples $N_s=10~000$. The open circles and open triangles correspond
to the CPU time of the Monte Carlo (MC) and MEC algorithms, respectively,
with one day as the unit. Generally speaking, the CPU time of the former is nearly
$100$ times larger than that of the latter, showing the great superiority
of the EMC algorithm.  As the lattice size $L$ increases, a power-law behavior
is observed for the CPU time, and the slope
$1.97$ of the EMC algorithm is a bit smaller than $2.25$
of the usual Monte Carlo algorithm which means that the EMC algorithm
is more efficient when the system size becomes larger,
pointing to a possibility of the large-scale simulations for the depinning
transition in DRFIM. Since the CPU time obeys $\delta t \sim L^2$, a
linear time complexity, i.e., $O(n)$, is revealed for the EMC algorithm,
further confirming that the algorithm is optimized.

\begin{figure}[tbp]
  \centering
   \includegraphics[clip, scale=0.3]{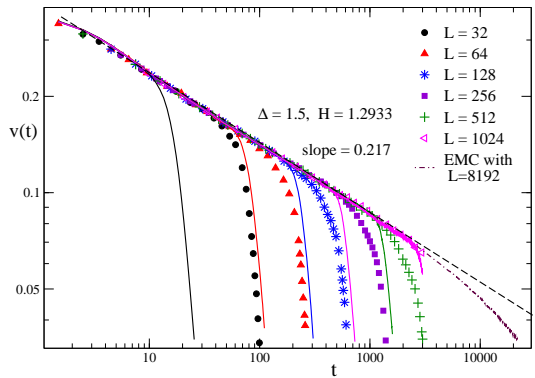}
   \caption{ Velocity of the domain interface $v(t)$ defined in Eq.~(\ref{equ140}) is
   displayed for different lattice sizes $L$ on a log-log scale. Other
   parameters, such as the driving field $H=1.2933$ and disorder
   strength $\Delta=1.5$, are used. The symbols represent the numerical results obtained
   by usual Monte Carlo method, and solid and dash-dotted line correspond to those obtained by EMC algorithm.
   Dashed line shows a power-law fit.}
   \label{f120}
\end{figure}

Besides, the accuracy of EMC algorithm are also be carefully
examined, in comparison with that of the usual Monte Carlo
algorithm. As shown in Fig.~\ref{f120}, the velocity of the domain
interface $v(t)$, as the order parameter of the depiinning
transition, is displayed as a function of the time $t$ for different
lattice sizes $L$. The symbols represent numerical results obtained
by the usual Monte Carlo method, and solid and dash-dotted lines
correspond to those obtained by the EMC algorithm. A driving field
$H=1.2933$ is then used, which was reported as the transition point
at the disorder strength $\Delta=1.5$ \cite{zho09,qin12}. With the
scaling form in Eq.~(\ref{equ180}), the dynamic behavior
$v(t) \sim t^{-\beta/\nu z}f(t/L^z)$ is deduced at $\tau=0$, and the finite-size
effect described by $f(t/L^z)$ can be easily controlled, i.e., it
rapidly approaches a constant as $L$ increases.

Two distinguishable scaling regimes are found in numerical results,
which are separated by a characteristic time scale $t_L \sim L^z$.
When $t < t_L$, the symbols of different lattice size $L$ nicely
collapse to a master curve $v(t)\sim t^{-\beta/\nu z}$ with the
slope $0.217(2)$, meaning an absence of the finite-size effect. The
symbols are overlap with the corresponding solid lines, confirming
that the numerical results obtained by these two algorithms are
almost the same. Though the deviations of the
symbols and lines are observed in the second time regime with $t>t_L$, one
can use a sufficiently large lattice size, e.g., $L=8192$, to make
sure all the results in this paper staying in the first regime with
$t_{max} < t_{L}$. Consequently, the EMC algorithm not only
significantly saves the computing cost, but also has a high accuracy.
Interestingly, the tail of the master curve, depicted by the
dash-dotted line, exhibits a significant deviation from the
power-law behavior marked by the dashed line, as shown in
Fig.~\ref{f120}. It indicates that the driving field $H=1.2933$ is
smaller than the exact transition point $H_c$.

\subsection{Depinning phase transition}

\begin{figure}[tbp]
  \centering
   \includegraphics[scale=0.3]{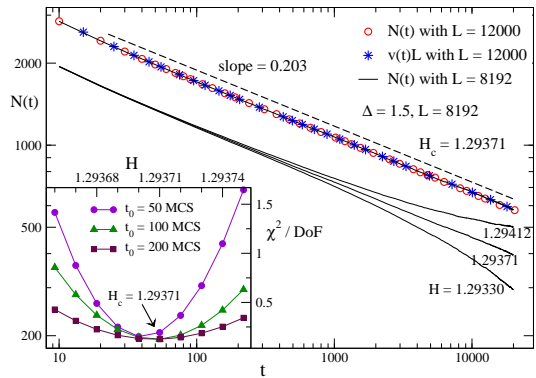}
   \caption{Dynamic relaxation of the activated-spin number $N(t)$ is plotted at
   $L=8192$ for different driving fields $H$ on a log-log scale. For clarity,
   the curve at the critical point $H_c=1.29371$ is shifted up, in comparison
   with the number $N(t)$ of the activated spin and the velocity $v(t)L$  of
   the domain interface at $L=12~000$. Dashed line shows a power-law fit.
   In the inset, the fitting error $\chi^2/{\rm DoF}$ is shown for different waiting
   time $t_0=50, 100$, and $200$ MCS. }
   \label{f130}
\end{figure}

Extensive numerical simulations are performed with the EMC
algorithm to identify the depinning transition in disordered media.
Unless otherwise stated, the lattice size $L=8192$ is used in the following.
Fig.~\ref{f130} shows the dynamic relaxation
of the number $N(t)$ of the activated spin under different driving fields
$H$. Similar with the velocity $v(t)$ of the domain interface, the
number $N(t)$ drops rapidly down for a small $H$,
while approaches a constant for a large $H$. Searching for the best
power-law behavior, one can locate the transition field
$H_c=1.29371(4)$, much more precise than the previous one
$1.2933(2)$ obtained with the usual Monte Carlo method \cite{zho09,qin12}.
In addition, another simulations with the lattice size $L'=12~000$
are also carried out. Rescaled by a factor $L'/L$, the solid line, i.e., $N(t)$ at
$L=8192$, is in perfect agreement with the open circles and stars,
corresponding to $N(t)$ and $v(t)L'$ at $L'=12~000$, respectively,
confirming that finite-size effect is already negligibly small.
Moreover, an almost-perfect power-law decay $N(t)=v(t)L\sim
t^{\beta/\nu z}$ is observed at the transition point $H_c=1.29371$.
A direct measurement from the slope of the fitting yields $\beta/\nu
z = 0.203(1)$.

\begin{figure}[tbp]
  \centering
   \includegraphics[scale=0.3]{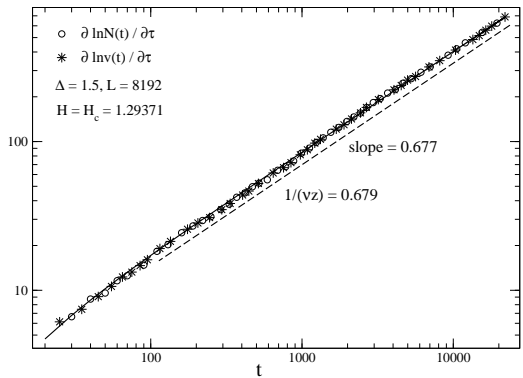}
   \caption{The logarithmic derivatives of $N(t)$ and $v(t)$ are displayed with
   open circles and stars, respectively, at the critical point $H_c=1.29371$
   for $\Delta=1.5$ and $L=8~192$. The dashed line represents a power-law fit,
   and the solid line shows a power-law fit with the correction defined in Eq.~(\ref{equ250}). }
    \label{f140}
\end{figure}

To obtain a more accurate value of the transition point, the error
of the power-law fitting $\chi^2/{\rm DoF}$ \cite{cha01} is carefully
examined within a very narrow $H$-regime $[1.29366, 1.29375]$, as shown
in the inset of Fig.~\ref{f130}.
The critical point can be determined by judging the location of the
minimization of $\chi^2/{\rm DoF}$. For different waiting time
$t_0=50, 100$ and $200$ MCS, an almost same value $H_c=1.29371(1)$
is obtained, further showing that our result is robust and precise.
In Fig.~\ref{f140}, the logarithmic derivatives of $N(t)$ and $v(t)$
in the neighborhood of $H_c=1.29371$ are shown with open circles and
stars, respectively. With the scaling form in Eq.~(\ref{equ210}),
the exponent $1/\nu z=0.677(3)$ is measured from the slope of the
dashed line. To extend the fitting to earlier times of the numerical
data, the correction to scaling is considered with the form in
Eq.~(\ref{equ250}). The exponent $1/\nu z=0.68(1)$ is then determined,
consistent with the previous one within the error bar.

\begin{figure}[tbp]
  \centering
   \includegraphics[scale=0.3]{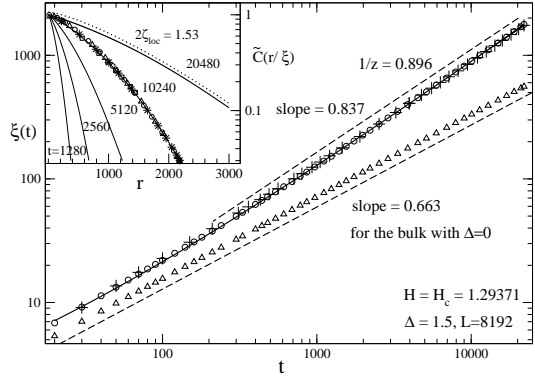}
   \caption{Time evolution of the nonequilibrium spatial correlation length $\xi(t)$ extracted from $C(r, t)$ is plotted
   for the domain interface with the disorder $\Delta=1.5$ (open circles) and the bulk with $\Delta=0$ (open triangles).
   Other parameters $H=1.29371$ and $L=8192$ are set. The pluses correspond to the function $F(t)$ defined in Eq.~(\ref{equ165}),
   dashed lines represent power-law fits, and solid line shows
   a power-law fit with the correction. In the inset, the scaling function $\widetilde{C}(r/\xi(t))$ is shown on a
   linear-log scale. Data collapse is demonstrated at $t = 10~240$ MCS, and the dotted line
   at $t=20~480$ represents a fit with the form in Eq.~(\ref{equ240}).
   }
   \label{f150}
\end{figure}

Besides the velocity of the domain interface, we also investigate
the roughness function $\omega^2(t)$ and the correlation function
$C(r,t)$ defined in Eqs.~(\ref{equ150}) and (\ref{equ160}),
respectively. With the scaling form of $C(r,t)$ in
Eq.~(\ref{equ220}), numerical data at different time $t$
collapse to the curve at $t'= 10~240$ MCS by rescaling $r$ to
$[\xi(t')/\xi(t)]r$ and C(r,t) to $[\omega^2(t')/\omega^2(t)]C(r,t
)$. Adopting this data-collapse technique \cite{zho14}, one can extract
the nonequilibrium correlation length $\xi(t)$ from the correlation function
$C(r,t)$. The results are shown in Fig.~\ref{f150} at the depinning field $H_c=1.29371$ for
the domain interface with $\Delta=1.5$ (open circles) and the
bulk with $\Delta=0$ (open triangles). Power-law behaviors are
observed for both of them with the slopes $1/z=0.837(5)$ and
$1/z_b=0.663(5)$, respectively. However, an obvious deviation from
the power law is found in the early times of the curve composed of
open circles, thereby correction to scaling should be considered.
With the correction form in Eq.~(\ref{equ250}), we refine the
value of the exponent $1/z=0.896(7)$. Moreover, a perfect
coincidence between the open circles and pluses are observed,
pointing to the relationship $F(t) \sim \xi(t)$, consistent with the
prediction in earlier work \cite{zho09}. In the inset, the scaling
function $\widetilde{C}(r/\xi(t))$ is plotted as the function of
$r$. Data of different time $t$ nicely collapse to the curve at
$t=10~240$ MCS by rescaling $r$ to $r/\xi(t)$, confirming the
accurate of the correlation length. Moreover, the exponent
$2\zeta_{loc}=1.53(4)$ is measured by the fitting with
Eq.~(\ref{equ240}).

With the correlation length $\xi(t)$ at hand, we then study the roughening process
of the domain interface at the depinning transition.
In Fig.~\ref{f150}, $\omega^2(t)$ is plotted as a function of
$\xi(t)$ on a log-log scale. Obviously, $\omega^2(t)$ at the
transition field $H_c=1.29371$ with $\Delta=1.5$ shows a cleaner
power-law behavior than $\xi(t)$ does, and the roughness exponent
$2\zeta=2.00(2)$ is measured based on Eq.~(\ref{equ230}). It is much
larger than $2\zeta_b=0.97(1)$ for the bulk with $\Delta=0$. In the
inset, the roughness function $\omega(t)$ is displayed with the solid line,
and the direct measurement of the slope yields $2\zeta/z=1.65(2)$.

\begin{figure}[tbp]
  \centering
   \includegraphics[scale=0.3]{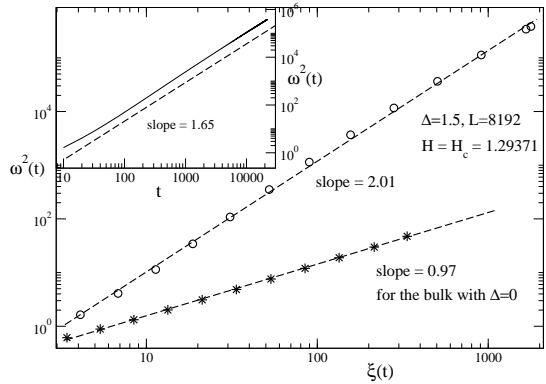}
   \caption{ The roughness function $\omega^2(t)$ against the correlation length $\xi(t)$ is displayed
   with open circles and stars, corresponding to the cases with the disorder strengthes $\Delta=1.5$ and $0$,
   respectively. In the inset, time evolution of $\omega^2(t)$ is shown for $\Delta=1.5$.
   Dashed lines represent power-law fits in Eq.~(\ref{equ250}). }
   \label{f160}
\end{figure}

Finally, we summarize all the measurements of the critical exponents
in Table.~\ref{t100}. The transition field $H_c=1.29371(1)$ is more
precise than the earlier result $1.2933(2)$ \cite{zho09,qin12}, and
significant deviations (reaching nearly $10\%\sim30\%$) are also
observed in the exponents $\nu, z$, and $\zeta$, showing the
necessity of large-scale simulations for the depinining.
For comparison, the exponents of the QEW equation
\cite{fer13, kim06} are also shown in the table with the same
lattice size $L=8~192$. The contention that the depinnning transition
of the DRFIM and QEW equation are not in a same universality class
is supported by the distinct differences of the exponents
between these two models, especially in $\beta, z$, and $\zeta$.
According to the arguments in Refs.\cite{zho09,zho10a}, the
difference is mainly due to the overhangs and islands created in the
depinning process of DRFIM.

\begin{table}[h]\centering
\caption{The transition point $H_c$ and critical exponents for the
depinning transition in DRFIM are obtained from the extensive
simulations with the EMC algorithm, in comparison with those
in the previous work and the QEW equation. }
\begin{tabular}[t]{l c c c c}
\hline
\hline  &   &  QEW  &  \multicolumn{2}{c}{DRFIM}  \\
\hline  &   &   Ref.\cite{fer13,kim06}   &  Previous work \cite{zho09,qin12} & This work  \\

\hline
$v(t)$        &   $H_c$            &            &      1.2933(2)      &  1.29371(1)    \\
              &   $\beta$          &   0.245(6) &      0.295(3)       &  0.299(3)      \\
              &   $\nu$            &   1.333(7) &      1.02(2)        &  1.32(2)      \\
$\xi(t)$      &   $z $             &   1.433(6) &      1.33(1)        &  1.12(1)       \\
$\omega^2(t)$ &   $\zeta$          &   1.250(5) &      1.14(1)        &  1.00(1)       \\
$C(r,t)$      &   $\zeta_{loc}$    &            &      0.735(8)       &  0.76(2)        \\
$H \gg H_c$   &   $z_b$            &            &      1.50(1)        &  1.51(1)        \\
              &   $\zeta_{b}$      &            &      0.49(1)        &  0.485(5)       \\
\hline \hline
\end{tabular}
\label{t100}
\end{table}

\subsection{New universality class}

\begin{figure}[bp]
  \centering
   \includegraphics[scale=0.3]{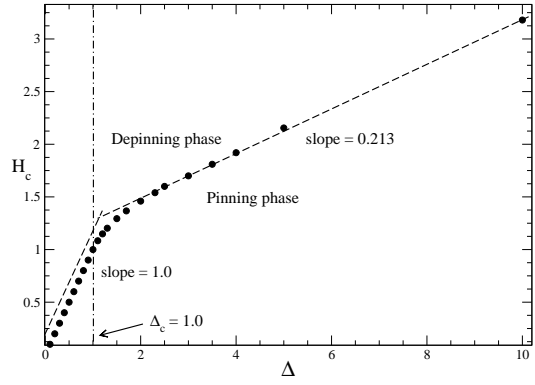}
   \caption{ The phase boundary of the pinning-depinning transition $H_c$ is plotted as a function of the disorder strength
   $\Delta$. The vertical dash-dotted line indicates a critical disorder $\Delta_c=1.0$, where the transition is of first order when $\Delta
   \leq \Delta_c$ and of second order when $\Delta > \Delta_c$. Dashed lines represent linear fits. }
   \label{f170}
\end{figure}

\begin{figure}[tbp]
  \centering
   \includegraphics[scale=0.3]{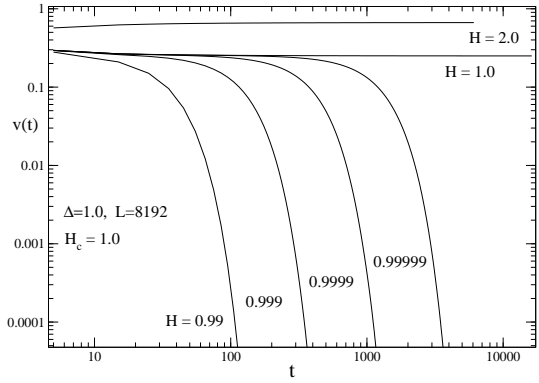}
   \caption{ The velocity of the domain interface $v(t)$ is plotted at the critical value of the disorder $\Delta_c=1.0$
   for different driving fields $H$. The velocity $v(t)$ drops to zero quickly even at $H=0.99999$, a slightly smaller than the transition point
   $H_c=1.0$ where $v(t)$ approaches a nonzero constant. }
   \label{f180}
\end{figure}

Does the depinning transition in the DRFIM belong to a universality class? To solve
this issue, comprehensive simulations have been performed for different strengthes of the disorder (varying from $\Delta=0$ to $10$) in DRFIM and
different types of the disorder, taking the random-bond disorder in the DBDIM as another example. As shown in Fig.~\ref{f170}, the phase boundary
is displayed separating the depinning phase from the pinning phase.  The vertical dash-dotted line indicates
a critical value $\Delta_c=1.0$ of the disorder, where the phase transition is of first order when $\Delta \leq \Delta_c$ and of second order when $\Delta > \Delta_c$.
In both of them, linear behaviors are observed with the slopes $1.00$ and $0.213(4)$, respectively. Unfortunately, the disorder strength
$\Delta=1.5$ which has been carefully investigated before is nearby the crossover between these two regimes,
suggesting that the values of the exponents in Table.~\ref{t100} may be $\Delta$-dependent.

\begin{figure}[tbp]
  \centering
   \includegraphics[scale=0.3]{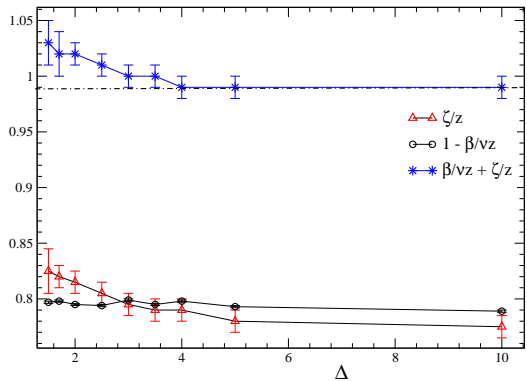}
   \caption{ Variation of the critical exponents $1-\beta/\nu z$, $\zeta/z$ and $\beta/\nu z + \zeta/z$ are displayed
   as a function of the disorder strength $\Delta$. The dash-dotted line points to an asymptotic
   value $\beta/\nu z + \zeta/z=0.99(1)$ in the regime of the strong disorder. }
   \label{f190}
\end{figure}

As an example, the velocity of the domain interface $v(t)$ at the critical point $\Delta_c=1.0$ is shown in Fig.~\ref{f180}. Quite different from
those in Figs.~\ref{f120} and \ref{f130}, $v(t)$ decays exponentially at $H \leq 0.99999$, and approaches a nonzero constant at $H \geq 1.00000$.
A huge jump is then found in the steady-state velocity as the driving field is changed by only a paltry amount of $10^{-5}$,
inferring that it is a typical first-order phase transition, and the transition point is $H_c=1.00000$.
Similar phenomena are also found for the disorder strength $\Delta < \Delta_c$. While power-law decays of the domain interface are found
at $H_c$ for $\Delta>\Delta_c$, corresponding to the second-ordered phase transition. The critical exponents with respect to the disorder strength $\Delta$
are investigated carefully, and a scaling relation $\beta/\nu z +\zeta/z=0.99(1)$ is revealed in Fig.~\ref{f190} for sufficiently strong disorder,
consistent with the prediction of the scaling theory $\beta=\nu(\zeta -z)$ in the QEW equation \cite{nat92, dou02}. It means that the roughness exponent $\zeta$
is not independent, and the roughening process of the domain interface $\omega^2(t) \sim t^{2\zeta/z}$ is merely induced by the depinning transition with
$h(t) \sim t^{1-\beta/\nu z}$. In contrast, the other scaling relation $\nu=1/(2-\zeta)$ is apparently violated, indicating the breaking
down of the statistical tilt symmetry.

In addition, the robustness of the exponents $\zeta/z$ and $1-\beta/\nu z$ is affirmed in Fig.~\ref{f190} for a wide window $\Delta \in [0, 10]$,
indicating a universality class. Furthermore, we summary the values of the exponents $\beta, \nu, z, \zeta, \zeta_{loc}$,
and $\beta/\nu z + \zeta/z$ in Table.~\ref{t110} for two cases with the disorder strengthes $\Delta=5.0$
and $10.0$, respectively. For each exponent, a perfect consistency between these two cases is found within the error bars.
In the same way, the depinning transition in the DBDIM is also investigated with a Gaussian distribution of the random-bond disorder.
The standard deviation $\sigma=1.5$ is used in the simulations, which is already sufficiently large for the depinning transition.
The results of these critical exponents listed in Table.~\ref{t110} are almost the same with those obtained from the DRFIM with $\Delta=5.0$ and $10.0$, but
significantly differ from those of the QEW equation shown in Table.~\ref{t100}. It further supports our conclusion that the depinning transition in the
Ising-type lattice models belong to a new universality class, though the scaling relation $\beta=\nu(z-\zeta)$ is still valid.

\begin{table}[h]\centering
\caption{The critical exponents of DRFIM and DBDIM are listed for the
depinning transition in the strong-disorder regime. The values of
these cases are consistent with each other, indicating a new universality
class irrelevant to the disorder. }

\begin{tabular}[t]{l c c c c c c}
\hline
\hline  Disorder &  $\beta$ &  $\nu$  & $z$ & $\zeta$  & $\zeta_{loc}$  &  $\beta/\nu z + \zeta/z$   \\
\hline
DRFIM  & & & & & & \\
$\Delta=5$         &    0.304(5)   &  1.35(3)   &   1.12(1)  &  0.90(1)    &  0.67(1)  & 0.99(1) \\
$\Delta=10$        &    0.301(5)   &  1.29(3)   &   1.12(1)  &  0.89(1)    &  0.64(1)  & 0.99(1)  \\
DBDIM  & & & & & & \\
$\sigma=1.5$         &    0.306(5)   &  1.31(3)   &   1.14(1)  &  0.90(1)    &  0.65(1)  & 1.00(1)  \\
\hline \hline
\end{tabular}
\label{t110}
\end{table}

\section{Conlusion}
The depinning transition in Ising-type lattice models have been systematically investigated with
the extensive simulations based on the developed EMC algorithm. In comparison with the usual Monte Carlo
method, the EMC algorithm shows great superiority in the efficiency of the simulations, i.e.,
only about a percent of the CPU time is needed. Taking the DRFIM with the disorder strength $\Delta=1.5$ as an example,
we determined the transition field and critical exponents based on the short-time scaling form.
The accuracy of the results is significantly improved by large-scale simulations with the lattice size up to $L=8~192$,
compared to those in earlier literature obtained at $L=1024$ \cite{zho09,qin12}. The phase diagram of the dpinning transition
is then identified with a critical value of the disorder strength $\Delta_c=1.0$, separating the first-order transition form the second-order one.
In the strong-disorder regime, the robustness of the exponents is uncovered for different strengthes (varying from $\Delta=0$ to $10$) and
different types of the disorder (random filed in the DRFIM and random bond in the DBDIM). The results
$\beta=0.304(5), \nu=1.32(3), z=1.12(1)$, and $\zeta=0.90(1)$ are quite different from those in the universality class of the Edwards-Wilkinson equation,
though the scaling relation $\beta=\nu(\zeta -z)$ remains. It indicates that the depinning transition in the Ising-type lattice models belongs to a
new universality class,  due to the intrinsic anomalous scaling and spatial multiscaling.  Moreover, the local roughness exponent $\zeta_{loc}=0.65(1)$
is then measured from the correlation function $C(r, t)$, comparable with the experimental results in the ultrathin Pt/Co/Pt films \cite{lem98}.

{\bf Acknowledgements:} This work was supported in part by the National Natural Science Foundation of China under Grant Nos.~$11205043$ and $11304072$, and
the authors thank Lu Wang and Tianci Zhou from the Zhejiang Universality for the insightful discussion.

\bibliographystyle{elsarticle-num}
\bibliography{domain,zheng}

\end{document}